\documentclass{article}[11pt]
\usepackage{a4wide,bm,bbm,epsfig,graphics,booktabs,amsmath,amssymb,amsfonts,color,epstopdf,dsfont}
\usepackage[
colorlinks=true,
linkcolor=black,
breaklinks=true,
urlcolor=blue,
citecolor=green]{hyperref}

\newcommand{\non}{\nonumber}
\newcommand{\al}{&\!\!\!}
\newcommand{\mO}{\mathcal{O}}
\newcommand{\order}[1]{\mathcal{O}\left(#1\right)}
\newcommand{\Lag}{\mathcal{L}}

\newcommand{\Tr}[1]{\left\langle #1 \right\rangle}
\newcommand{\mM}{\mathcal{M}}
\newcommand{\unitmatrix}{\mathds{1}}
\newcommand{\intd}{i\mu^{4-d}\!\!\int\!\!\frac{d^dl}{(2\pi)^d}}

\begin{document}
\title{$\theta$-dependence of the lightest meson resonances in QCD}

\author{Neramballi Ripunjay Acharya$^{1,}$\footnote{Email address:
      \texttt{acharya@hiskp.uni-bonn.de} },~
      Feng-Kun Guo$^{1,}$\footnote{Email address:
      \texttt{fkguo@hiskp.uni-bonn.de} },~
      Maxim Mai$^{1,}$\footnote{Email address:
      \texttt{mai@hiskp.uni-bonn.de} },~
      Ulf-G. Mei{\ss}ner$^{1,2,}$\footnote{Email address:
      \texttt{meissner@hiskp.uni-bonn.de} }\\[0.5em]
      {\it\small$^1$Helmholtz-Institut f\"ur Strahlen- und Kernphysik and Bethe
      Center for Theoretical Physics,}\\
      {\it\small Universit\"at Bonn, D-53115 Bonn, Germany}\\
      {\it\small$^2$Institute for Advanced Simulation, Institut f\"{u}r
       Kernphysik, J\"ulich Center for Hadron Physics,}\\
      {\it\small and JARA-FAME, Forschungszentrum J\"{u}lich, D-52425 J\"{u}lich, Germany}}

\maketitle
\begin{abstract}

We derive the pion mass and the elastic pion-pion scattering amplitude in the 
QCD $\theta$-vacuum up to next-to-leading order in chiral perturbation theory. 
Using the modified inverse amplitude method, we study the $\theta$-dependence 
of the mass and width of the light scalar  meson $\sigma(500)$ and the vector meson $\rho(770)$.

\end{abstract}

\section{Introduction}

Light flavor quantum chromodynamics (QCD) with up, down and strange quarks is a
fascinating theory, that features only a few parameters. Variations of these parameters
have been explored in great detail in the last decade. We mention in
particular varying the quark masses to make contact with lattice
simulations, but also variants of the theory with more light quark
flavors or different numbers of colors have been studied. These studies
mostly focused on the fate and exploration of the dynamical and explicit
symmetry breaking that light flavor QCD exhibits
 A much less explored territory
relates to the $\theta$-term of QCD,
\begin{equation}
{\cal L}_\theta = -\displaystyle\frac{\theta}{64 \pi^2} \,
\epsilon^{\mu\nu\rho\sigma} G^a_{\mu\nu}G^a_{\rho\sigma}\,,
\end{equation}
where $a =1, \ldots,8$ are color indices and $G_{\mu\nu}$ is the gluon field strength
tensor. While the upper limit on the neutron electric dipole moment poses
a stringent limit on the value on $\theta$, where the latest determination gives
$|\theta| < 7.6 \cdot 10^{-11}$~\cite{Guo:2015tla}, it has been argued that based on the
string theory landscape, $\theta$ might take natural values and that it is difficult
to achieve a tiny value for it, see e.g. Ref.~\cite{Banks:2003es}.
Independently of that, it is interesting to explore QCD at larger
values of $\theta$. QCD at $\theta \sim \pi$ was already investigated in
Refs.~\cite{Smilga:1998dh,Tytgat:1999yx}. Also, the pion and the nucleon mass
were calculated at fixed topology in Ref.~\cite{Brower:2003yx} to explore the
connection between lattice QCD and physical observables. Further,
Ubaldi~ \cite{Ubaldi:2008nf}  studied the effects that a non-zero
strong-CP-violating parameter  would have on the deuteron
and di-proton binding energies and on the triple-alpha process using a somewhat
simplified nuclear modeling.  Even so the relevant
energy scales in these systems exhibit some fine-tuning, no dramatic effect of varying
$\theta$ was found. We also note a recent study on the
relation between confinement and the $\theta$-vacuum, see Ref.~\cite{Kharzeev:2015xsa}.

The $\theta$-dependence of the pion mass has been given at leading order (LO) in
chiral perturbation theory (ChPT)~\cite{Brower:2003yx}. Here, we want to derive
the pion mass in the $\theta$-vacuum up to next-to-leading order (NLO) as well 
as the corresponding pion-pion ($\pi\pi$) elastic  scattering amplitude. Furthermore, we will 
calculate the
$\theta$-dependence of the lightest resonances in QCD, the scalar meson
$\sigma(500)$ and the vector meson $\rho(770)$.
These are not only interesting by themselves, but also are important components
in precision modelings of the nuclear forces, as used e.g. in the work on
extracting limits on variations of the Higgs vacuum expectation value 
from the element abundances in Big Bang nucleosynthesis \cite{Berengut:2013nh} 
(for related works using different frameworks, see Refs.~\cite{Damour:2007uv,Hall:2014dfa}). 
In the following, we assume that the $\sigma$ and the $\rho$ are generated from a
resummation of pion-pion interactions evaluated at NLO in ChPT
 in the $\theta$-vacuum. Having calculated the
$\theta$-dependent pion-pion scattering amplitude, it is straightforward to
implement it into a unitarization (resummation) scheme that can generate the
light resonances and thus gives their $\theta$-dependent properties.
To be specific, we make use of the so-called (modified) inverse amplitude method,
which is one, but not the only available, unitarization scheme that allows to
perform this task.

Our work is organized as follows. In Sec.~\ref{sec:L} we discuss the
chiral effective Lagrangian in the $\theta$-vacuum, with particular
emphasis on the two-flavor formulation and also including strong isospin
breaking. Next, the $\theta$-dependent $\pi\pi$ scattering amplitude
is constructed at NLO in Sec.~\ref{sec:pipi}. Armed with that, we then
come to the central section~\ref{sec:rhosigma}, where the mass and the
width of the $\sigma$ and the $\rho$ are calculated as a function of $\theta$.
We end with a short summary and outlook in Sec.~\ref{sec:sum}. The appendix
contains some discussion of the vacuum alignment at NLO.

\section{Chiral effective Lagrangian in the $\theta$-vacuum}
\label{sec:L}

At the lowest order, $\order{p^2}$, the SU($N$) chiral Lagrangian in the
$\theta$ vacuum is~\cite{Gasser:1983yg,Gasser:1984gg}
\begin{equation}
\Lag_2 = \frac{F^2}{4} \Tr{D_\mu U^\dag D^\mu U}+
\frac{F^2}{4} \Tr{ \chi U^\dag + \chi^\dag U},
\end{equation}
where $F$ is the pion decay constant in the chiral limit and $\chi = 2 B
\mM\,\exp(i\theta/N)$. Here, $\mM$ is the real and diagonal quark mass
matrix, and the low-energy constant (LEC) $B=\Sigma/F^2$,
with $\Sigma$ the absolute value of the flavor-averaged quark condensate
in the chiral limit. The field $U\in$ SU($N$) collects the Goldstone bosons of the theory. 
However, only for $\theta=0$ the vacuum expectation value of it, which is the solution of 
the equations of motion for the zero-momentum mode, is trivial, i.e. $U_0 = \unitmatrix$.
In the general case, the vacuum is shifted from the unit matrix, and the vacuum alignment
can be determined by minimizing the potential energy. Therefore, it is useful to separate
the ground state $U_0$ from the quantum fluctuation  $\tilde U$ containing the Goldstone
boson fields as $U(x)=U_0\,\tilde{U}(x)$. Thus, the ground state of the theory is given by
minimizing the potential energy
\begin{equation}
 V_2 = - \frac{\Sigma}{2} \Tr{ \left( U_0^\dag\, e^{i\theta/N} + U_0^{}\,
  e^{-i\theta/N} \right) \mM }\,.
\end{equation}
Because $\mM$ is diagonal, $U_0$ can be taken as diagonal as well without loss of 
generality. The special unitary matrix $U_0$ can be parametrized as
\begin{equation}
  U_0 = \text{diag} \left\{ e^{i\varphi_1}, e^{i\varphi_2}, \ldots,
  e^{i\varphi_N} \right\},\qquad \sum_f \varphi_f = 2 n \pi~(n\in\mathbb{Z}) \,.
\end{equation}
This leads to
\begin{equation}
  V_2 = - \Sigma\, \text{Re} \Tr{ e^{-i\theta/N}U_0\,\mM} = -\Sigma \sum_f
  \cos\left( \varphi_f - \frac{\theta}{N} \right) m_f \, ,
\end{equation}
with $m_f$ the quark mass of flavor $f$.

For $\theta=\pi$, because the theory is periodic in $\theta$ with a period $2\pi$, 
the Lagrangian is invariant under CP and P transformations as they change $\theta$ to
$-\theta$. However, it is well-known that at $\theta=\pi$, there is a
spontaneous CP breaking, called Dashen's phenomenon, because the
CP conserving stationary point of the action is in fact a maximum and there are
two degenerate CP violating vacua which are obtained by minimizing the potential
energy \cite{Dashen:1970et}.
\subsection{Two-flavor case without isospin symmetry}

In the rest of the paper, we will consider the two-flavor case and use the following parametrization
\begin{equation}
U_0 = {\rm diag}\{e^{i\varphi}, e^{-i\varphi}\}, \qquad \tilde U = e^{i \sqrt{2}\Phi/F},
\qquad \Phi = \frac1{\sqrt{2}}
\begin{pmatrix}
  \pi^0 & \sqrt{2}\pi^+ \\
 \sqrt{2} \pi^- & - \pi^0
\end{pmatrix}\,.
\end{equation}
We see that the angle $\varphi$ and the neutral pion field always appear in a
linear combination $\varphi + \pi^0/F$. Therefore, finding the stationary
solution for $U_0$ by minimizing the potential energy with respect to $\varphi$
is equivalent to removing the tree-level tadpole for the neutral
pion~\cite{Dashen:1970et}.  The minimization of $V_2$ gives~\cite{Brower:2003yx}
\begin{align}
(m_u + m_d) \sin\varphi \cos\frac{\theta}{2} &- (m_u - m_d) \cos\varphi
\sin\frac{\theta}{2} = 0 \non\\
\Leftrightarrow~\tan \varphi &= - \epsilon \tan\frac{\theta}{2}\,,
\label{eq:phi}
\end{align}
where the average light quark mass $\bar m = (m_u+m_d)/2$ and the parameter
$\epsilon = (m_d - m_u)/(2\bar m)$, that quantifies strong isospin breaking,
are introduced. Actually, at $\theta=\pi$ and $\epsilon=0$ the Eqs.~\eqref{eq:phi} do not depend on $\varphi$ at all. This leads to a paradoxical situation of continuous 
vacuum degeneracy discussed in \cite{Smilga:1998dh,Tytgat:1999yx}
and resolved in \cite{Creutz} taking into account terms of the NLO chiral Lagrangian.

In the present work we also consider the NLO chiral Lagrangian~\cite{Gasser:1983yg}. 
In the SU(2)$\times$SU(2) notation of e.g. Ref.~\cite{Bellucci:1994eb} it reads
\begin{eqnarray}
  \Lag_4 \al=\al \frac{l_1}{4} \Tr{D_\mu U^\dag D^\mu U}^2 + \frac{l_2}{4}
  \Tr{D_\mu U^\dag D_\nu U} \Tr{D^\mu U^\dag D^\nu U} + \frac{l_3}{16} \Tr{
  \chi^\dag U + \chi U^\dag }^2 \nonumber\\
  \al\al + \frac{l_4}{4}
  \Tr{ D_\mu \chi^\dag D^\mu U + D_\mu \chi D^\mu U^\dag } +
  \frac{l_5}{4}\Tr{U^\dag F_{\mu\nu}^R U F^{L,\mu\nu} } \nonumber\\
  \al\al + \frac{i\,l_6}{2} \Tr{ F_{\mu\nu}^R D^\mu U D^\nu U^\dag +
  F_{\mu\nu}^L D^\mu U^\dag D^\nu U } - \frac{l_7}{16} \Tr{\chi^\dag U - \chi
  U^\dag }^2 \nonumber\\
  \al\al + \frac{h_1+h_3}{4} \Tr{\chi^\dag \chi} + \frac{h_1-h_3}{2}
  \text{Re}\,(\text{det}\chi) - h_2 \Tr{F_{\mu\nu}^L F^{L,\mu\nu} +
  F_{\mu\nu}^R F^{R,\mu\nu}}\,,
  \label{eq:L4}
\end{eqnarray}
where
\begin{eqnarray}
D_\mu \mO \al=\al \partial_\mu \mO - i r_\mu \mO + i \mO l_\mu
~~\text{for}~~\mO = U,\,\chi  \nonumber\\
F_R^{\mu\nu} \al=\al \partial^\mu r^\nu - \partial^\nu r^\mu - i [r^\mu,r^\nu]\,,
\qquad
F_L^{\mu\nu} = \partial^\mu l^\nu - \partial^\nu l^\mu - i [l^\mu,l^\nu]
\end{eqnarray}
with $r^\mu$ and $l^\nu$ the right-handed and left-handed external fields,
respectively.

In principle, with the introduction of the NLO Lagrangian as well as the
one-loop contribution, the vacuum energy has changed, and the vacuum alignment
needs to be re-determined. In particular, the $l_7$ term in the NLO Lagrangian
is not minimized by the LO solution given in Eq.~\eqref{eq:phi}. This means that
the $l_7$ term induces a shift to the LO vacuum alignment. However, as shown in
Appendix~\ref{app:alignment}, this shift does not affect the calculation of the
pion masses and the $\pi\pi$ scattering amplitudes up to NLO. Therefore, it is
sufficient to consider the LO vacuum alignment in Eq.~\eqref{eq:phi} for our
purpose\footnote{We will not discuss
the complications at $\theta=\pi$. In that case, one needs to include
the $l_7$ term as it determines the whole dynamics~\cite{Smilga:1998dh}.}.

\subsection{$\theta$-dependence of the pion mass}

Substituting $U_0$ with $\varphi$ given by Eq.~\eqref{eq:phi} into the LO
Lagrangian, we get the LO pion mass squared in the
$\theta$-vacuum~\cite{Brower:2003yx}
\begin{equation}
 \mathring{M}^2(\theta) = 2 B \bar m \cos
\frac{\theta}{2} \, \sqrt{1 + \epsilon^2
 \tan^2\frac{\theta}{2}} \, ,
 \label{eq:mpiLO}
\end{equation}
which is the same for the neutral and charged pions.

At NLO, the pion masses receive contributions from both one-loop diagrams and
the $l_3$ and $l_7$ terms. The divergence in the one-loop diagrams cancel
exactly with that from $l_3$. We obtain
\begin{eqnarray}
  M_{\pi^+}^2(\theta) \al=\al  \mathring{M}^2(\theta) +
  \frac{\mathring{M}^4(\theta)}{F^2} \left( \frac1{32\pi^2} \ln
  \frac{\mathring{M}^2(\theta)}{\mu^2} + 2 l_3^r + 2 l_7
  \left(\frac{(1-\epsilon^2) \tan(\theta/2)}{ 1 + \epsilon^2\tan^2(\theta/2) }
  \right)^2 \right) , \nonumber\\
   M_{\pi^0}^2(\theta) \al=\al M_{\pi^+}^2(\theta) - 2l_7\,\frac{ \mathring{M}^4(\theta)}{F^2}
   \frac{ \epsilon^2 }{\cos^4({\theta}/{2}) \left(1 + \epsilon^2\tan^2({\theta}/{2}) \right)^2 } \, ,
   \label{eq:mpiNLO}
\end{eqnarray}
where $l_3^r$ is the scale-dependent finite part of $l_3$. At $\theta=0$, these 
expressions reduce to the standard SU(2)  relations derived in 
 Ref.~\cite{Gasser:1983yg}. Using the positivity bound for $l_7$ obtained in 
Ref.~\cite{Smilga:1998dh}, we find that the charged pion is always heavier than the neutral
one.

For easy reference, we give the corresponding formulae for the
much simpler isospin symmetric case with $m_u=m_d=\bar m$. In this case, the
stationary solution of the vacuum energy has $\varphi=0$. The pion mass up to
NLO in the $\theta$-vacuum has one additional term compared with that in the
$\theta=0$ case, and is given by
\begin{equation}
  M_\pi^2(\theta) =
  {M}^2(\theta) +
  \frac{{M}^4(\theta)}{F^2} \left( \frac1{32\pi^2} \ln
  \frac{{M}^2(\theta)}{\mu^2} + 2 l_3^r + 2 l_7
  \tan^2\frac{\theta}{2} \right) 
\end{equation}
with isospin symmetric LO pion mass
\begin{equation}
M_\pi^2(\theta) = 2 B \bar m\,\cos\frac{\theta}{2}\,. \label{eq:LOmass}
\end{equation}
One sees that even in the isospin symmetric case, the NLO pion mass depends on
$l_7$, and this additional term vanishes at $\theta=0$.

\section{$\pi\pi$ scattering amplitudes in a $\theta$-vacuum}\label{sec:pipi}

The $\pi\pi$ scattering amplitude at NLO is the building block to generate
the light mesons $\sigma(500)$ and $\rho(770)$ via unitarization. To be specific, we calculate the amplitude $A(s,t,u)=A_{\pi^+\pi^-\to \pi^0\pi^0}(s,t,u)$ which is used to get the following
combinations with definite isospin ($I = 0,1,2$)
\begin{eqnarray}
  T^0(s,t) \al=\al 3 A(s,t,u) + A(t,u,s) + A(u,s,t)\,,\non\\
  T^1(s,t) \al=\al A(t,u,s) - A(u,s,t)\,, \\
  T^2(s,t) \al=\al A(t,u,s) + A(u,s,t)\,.\non
\end{eqnarray}
Later, we will also need the partial-wave projection for given isospin $I$
and angular momentum $L$
\begin{equation}
 T_L^I(s) = \frac1{32\pi} \frac12 \int_{-1}^{+1} dz\, T^I(s,t)
 P_L(z)
\end{equation}
with $P_L(z)$ the pertinent Legendre polynomials.

\begin{figure}[tbh]
\centering
\includegraphics[width=\linewidth]{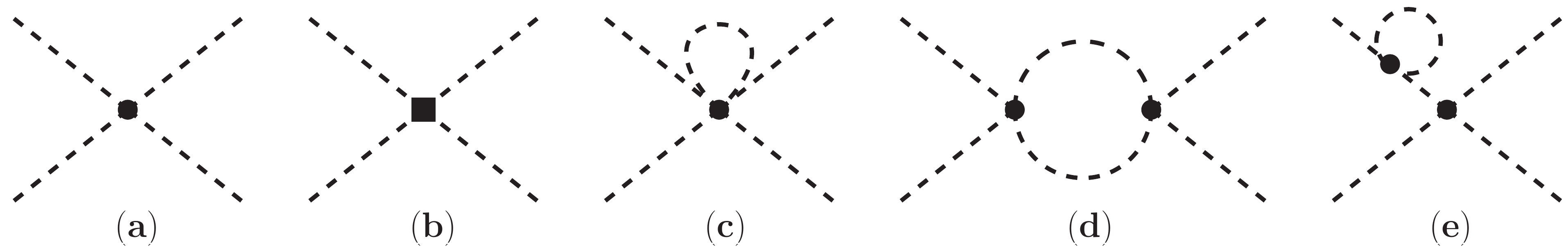}
\caption{Feynman diagrams for the $\pi\pi$ scattering amplitude up to
$\order{p^4}$. Here the filled circles and square denote the vertices from the
LO and NLO Lagrangian, respectively. The $t$ and $u$ channel two-point loops
are not shown.\label{fig:feyndiag}}
\end{figure}

Up to the order $\order{p^4}$, there are several contributions to the $\pi\pi$
scattering amplitude, as shown in Fig.~\ref{fig:feyndiag}. Diagram (a) gives
\begin{equation}
 A_\text{(a)}(s,t,u) = \frac1{3 F^2} \left\{3s + \mathring{M}^2_\theta-2
 \left[M_{\pi ^0}^2(\theta) +M_{\pi ^+}^2(\theta) \right] \right\} ,
 \label{eq:Aa}
\end{equation}
in terms of  $\mathring M_\theta^2\equiv\mathring{M}^2(\theta)$,
$M_{\pi^+}^2(\theta)$ and $M_{\pi^0}^2(\theta)$ given in Eqs.~\eqref{eq:mpiLO} and \eqref{eq:mpiNLO},
respectively. Diagram~(b) gives
\begin{eqnarray}
  A_\text{(b)}(s,t,u) \al=\al \frac{2\, l_1 }{F^4}\left(s-2
  \mathring M_\theta^2\right)^2 + \frac{l_2}{F^4} \left[4
 \mathring M_\theta^2  \left(s-2 \mathring M_\theta^2 \right) +t^2+u^2
  \right] +\frac{8\, l_3}{3 F^4}\mathring M_\theta^4 \nonumber\\
  \al\al +  \frac{32\, l_7 B^4 \bar m^4}{3 F^4} \left[ (1-\epsilon^2)^2
  \sin^2\theta - 2 \epsilon^2 \right] .
  \label{eq:Ab}
\end{eqnarray}
Diagram~(c) includes the tadpole vertex correction from both the neutral and
charged pions, and its contribution is
\begin{equation}
  A_\text{(c)}(s,t,u) = \frac1{18 F^4}\left(31\mathring M_\theta^2-20 s\right)
  A_0\left(\mathring M_\theta^2\right)~,
\end{equation}
$A_0(m^2)$ is the one-point loop integral (the tadpole) in $d$ space-time dimensions
\begin{equation}
  A_0(m^2) := \intd \frac1{l^2-m^2} \, ,
\end{equation}
with $\mu$ the scale of dimensional regularization.

The two-point loops, diagram (d) and the corresponding $t$- and $u$-channel crossed diagrams, give
\begin{eqnarray}
 A_\text{(d)}(s,t,u) \al=\al \frac1{18 F^4} \left[9 \left(
 \mathring M_\theta^4-s^2\right) B_0\left(s,
 \mathring M_\theta^2,\mathring M_\theta^2\right)+ 10 s\,
 A_0\left( \mathring M_\theta^2\right) \right] \nonumber\\
 \al\al + \frac1{6F^4} \Bigg\{ \left[2 \mathring M_{\theta }^4+2
 \mathring M_{\theta }^2 (t-2 u)+t (u-t)\right] B_0\left(t,\mathring M_{\theta
 }^2, \mathring M_{\theta }^2\right) +\frac23 (t-3 u)
 A_0\left(\mathring M_{\theta }^2\right)  \nonumber\\
 \al\al + \frac1{48 \pi ^2} \left(s- u\right)\left(t-6
 \mathring M_{\theta }^2\right) \Bigg\} \nonumber\\
\al\al + \frac1{6F^4} \Bigg\{ \left[2 \mathring M_{\theta }^4+2
 \mathring M_{\theta }^2 (u-2 t)+u (t-u)\right] B_0\left(u,\mathring M_{\theta
 }^2, \mathring M_{\theta }^2\right) +\frac23 (u-3 t)
 A_0\left(\mathring M_{\theta }^2\right)  \nonumber\\
 \al\al + \frac1{48 \pi ^2} \left(s- t\right)\left(u-6
 \mathring M_{\theta }^2\right) \Bigg\}\, ,
\end{eqnarray}
where the first line corresponds to the $s$-channel charged and neutral pion
loops, the second and third lines correspond to the $t$-channel loop, and the
last two lines are for the $u$-channel loop.
Here $B_0$ is the scalar two-point loop integral
\begin{equation}
  B_0(q^2,m_1^2,m_2^2) := \intd \frac1{(l^2-m_1^2)[(l+q)^2-m_2^2]}\, .
\end{equation}

We also need to take into account the wave function renormalization for all external lines which is represented by diagram (e). This amounts to
\begin{equation}
  A_\text{(e)}(s,t,u) = \frac12 \left(2\,\delta Z_{\pi^+} + 2\,\delta
  Z_{\pi^0}\right) \frac1{F^2} \left(s - \mathring{M}^2_\theta \right)
  = \frac4{3F^4} A_0\left(\mathring M^2_\theta\right) \left(s -
  \mathring{M}^2_\theta \right) ,
\end{equation}
where $\delta Z_\pi = Z_\pi -1$, with the wave function renormalization constant
for both the neutral and charged pions given by
\begin{equation}
Z_\pi = 1 + \frac2{3F^2} A_0\left(\mathring M^2_\theta\right) .
\end{equation}

Using dimensional regularization for the loop integrals and summing up all contributions, we obtain a UV divergence-free and scale-independent amplitude. The $l_3$ and $l_7$ terms in Eq.~\eqref{eq:Ab}
cancel with the same terms in Eq.~\eqref{eq:Aa} that enter through the NLO mass
expressions for the neutral and charged pions. The full amplitude reads
\begin{eqnarray}
\label{eq:Apipi}
  A(s,t,u) \al = \al \frac{s-\mathring M^2_\theta}{F^2} +  B(s,t,u) +
  C(s,t,u)\,, \\
  B(s,t,u) \al=\al \frac1{6 F^4} \Big\{ 3 (s^2-\mathring M^4_\theta)\bar J(s) +
  \left[ t(t-u) - 2 \mathring M_\theta^2\, t + 4 \mathring M_\theta^2\, u - 2
  \mathring M_\theta^4 \right] \bar J(t) \nonumber\\
  \al\al + \left[ u(u-t) - 2 \mathring M_\theta^2\, u + 4 \mathring M_\theta^2\,
  t - 2 \mathring M_\theta^4 \right] \bar J(u) \Big\}\,,\nonumber\\
  C(s,t,u) \al=\al \frac1{96 \pi^2 F^4} \left\{ 2 \left(\bar
  l_{1\theta}-\frac43\right) \left( s - 2 \mathring M_\theta^2 \right)^2 +
  \left( \bar l_{2\theta} - \frac56\right) \left[s^2 + (t-u)^2 \right] - 12
  \mathring M_\theta^2 s + 15 \mathring M_\theta^4 \right\}\,.
  \nonumber
  \label{eq:Amp}
\end{eqnarray}
Here, the $\bar l_{i\theta}$ are the scale-independent but quark mass-dependent,
and thus $\theta$-dependent, LECs which are related to the renormalized ones as
\begin{equation}
  l_i^r = \frac{\gamma_i}{32\pi^2} \left( \bar l_{i\theta} + \ln
  \frac{\mathring M_\theta^2}{\mu^2} \right), \qquad \gamma_1 = \frac13\, ,
  \quad \gamma_2 = \frac23\,.
\end{equation}
The finite loop function $\bar J$ is given by
\begin{equation}
  \bar J(s) = \frac1{16\pi^2} \left(\sigma(s) \ln \frac{\sigma(s)-1}{\sigma(s)
  +1} +2 \right)\,, \qquad \sigma(s) = \sqrt{ 1 - \frac{4 \mathring
  M_\theta^2}{q^2} }\,.
\end{equation}
We find that the $\pi\pi$ scattering amplitude up to NLO in a $\theta$-vacuum,
Eq.~\eqref{eq:Apipi}, takes exactly the same form as the well-known one in the
vacuum with $\theta=0$~\cite{Gasser:1983yg}, and the only change is to replace
everywhere $\mathring M^2(0)=2B\bar m$ by $\mathring M_\theta^2$ given by
Eq.~\eqref{eq:mpiLO}. The reason is that vertices from terms of the form
$\Tr{\chi^\dag U+\chi U^\dag}$ can always be written in terms of $\mathring
M_\theta^2$, while the $l_7$ term from diagram~(b) gets cancelled with the one
from diagram~(a). Such a property does not hold at higher orders. For instance,
considering the $\pi\pi$ scattering at $\order{p^6}$, there can be a one-pion
exchange diagram with two CP-violating three-pion vertices (see
Appendix~\ref{app:alignment}), which does not have any correspondence at
$\theta=0$. This behaviour of the $\pi\pi$ scattering amplitude is reminiscent
of the Kaplan-Manohar transformation \cite{Kaplan:1986ru}, 
which is an accidental symmetry of the chiral Lagrangian at NLO.

\section{$\theta$-dependence of the $\sigma$ and $\rho$ in the isospin limit}
\label{sec:rhosigma}

The $\sigma(500)$ and the $\rho(770)$ are the lightest two-flavor
non-Goldstone mesons. They can be obtained from the chiral perturbation
theory amplitudes by unitarization. There are various such unitarization
schemes on the market, like the inverse amplitude method (IAM)
to be used here~\cite{Dobado:1989qm}. In most cases, such a unitarization procedure
amounts to a resummation  of a certain class of diagrams to ensure exact two-body
unitarity, which is only perturbative in ChPT, but such resummations are usually
at odds with crossing symmetry. We do not want to enter a more detailed discussion
on these issues here (see e.g. the early work in Ref.~\cite{Gasser:1990bv}),
but rather employ the IAM as a tool to generate the light mesons
from the $\theta$-dependent pion-pion interaction, which automatically leads
to $\theta$-dependent properties of the $\sigma$ and the $\rho$.

The scattering amplitude for a given channel (with fixed isospin and angular
momentum) up to NLO in the IAM is given by
\begin{equation}
  T(s) = \frac{\left(T_{(2)}(s)\right)^2}{ T_{(2)}(s) - T_{(4)}(s) }\,,
\end{equation}
where $T_{(2)}(s)$ and $T_{(4)}(s)$ are the $\pi\pi$ scattering amplitudes of leading 
and next-to-leading chiral order. This form is valid in the channel with $I=J=1$ 
pertinent to the $\rho$-meson. As pointed out e.g. in Ref.~\cite{Hannah:1997sm},
it requires modification  in the $I=J=0$ channel due to the presence of
Adler zeros in the $S$-wave. The associated unphysical poles can be 
cancelled in rather natural way as derived in Ref.~\cite{GomezNicola:2007qj}
which is called the modified inverse amplitude method (mIAM). The corresponding 
scattering amplitude reads
\begin{eqnarray}
  T(s) \al=\al \frac{\left(T_{(2)}(s)\right)^2}{ T_{(2)}(s) - T_{(4)}(s) + A^{\rm mIAM}(s)
  }\,,\\
  A^{\rm mIAM}(s) \al=\al T_{(4)}(s_2) -
  \frac{(s_2-s_A)(s-s_2)\left(T_{(2)}'(s_2)-T_{(4)}'(s_4)\right)}{s-s_A}\,, \nonumber
\end{eqnarray}
where $s_A$ denotes the Adler zero of the full partial wave defined by the condition
$T(s_A)= 0$,  and $^\prime$ denotes differentiation with respect to $s$. 
The approximative Adler zeros at LO and NLO correspond to the energies
$s_2$ and $s_2+s_4$, determined by $T_{(2)} (s_2) = 0$ and $T_{(2)} (s_2+s_4) + T_{(4)} (s_2+s_4) = 0$, respectively. This mIAM has been used e.g. in Ref.~\cite{Hanhart:2008mx} to study
the quark mass dependence of the sigma and the rho. In particular, we will use the
LECs $l_1^r$ and $l_2^r$ as determined in that paper at the scale $\mu=770$~MeV,
\begin{equation}
  l_1^r = (-3.7\pm0.2)\times10^{-3},\qquad l_2^r = (5.0\pm0.4)\times10^{-3}\,.
\end{equation}
As mentioned in Ref.~\cite{Hanhart:2008mx}, the results of the IAM is
insensitive to the values of $l_3^r$ and $l_4^r$ as long as they are within the
uncertainties: $l_3^r = (0.8\pm3.8)\times10^{-3}$ and $l_4^r =
(6.2\pm5.7)\times10^{-3}$. Taking the central values and the measured values of
the pion mass and decay constant {$M_\pi=138.04$~MeV} (isospin averaged) and
$F_\pi=92.21$~MeV, the standard ChPT one-loop expressions yield {${M=139.46}$~MeV} and
$F=86.43$~MeV.

The masses and widths of the $\rho$ and $\sigma$ resonances can be obtained by
searching for the poles in the complex $s$-plane of the unitarized amplitudes.
When the resonances are above the $\pi\pi$ threshold, which is the case for the
physical pion mass, we need to search for poles in the second Riemann sheet. 
The corresponding pole positions read
\begin{eqnarray}
  \sqrt{s_\sigma} = (443.1 - i\,217.4)~\text{MeV}\,, \qquad
  \sqrt{s_\rho} =   (751.9 - i\,75.4)~\text{MeV}\,.
\end{eqnarray}
We note that the mass of the $\rho$ comes out somewhat below the physical value
as it is common in such unitarization procedures. For a more detailed discussion
on this issue, see e.g. Refs.~\cite{Oller:1998hw,Guo:2012ym}.
\begin{figure}[t]
\centering
\includegraphics[width=0.49\linewidth]{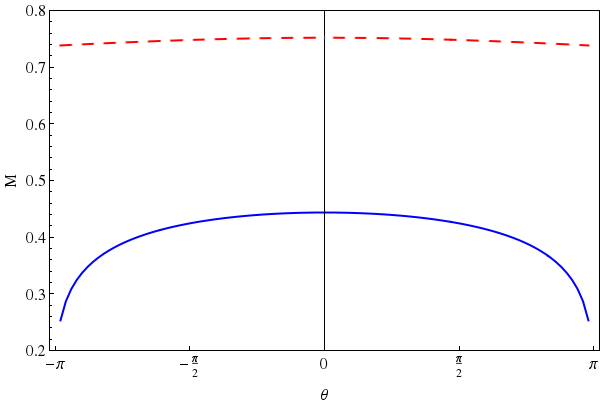}
\includegraphics[width=0.49\linewidth]{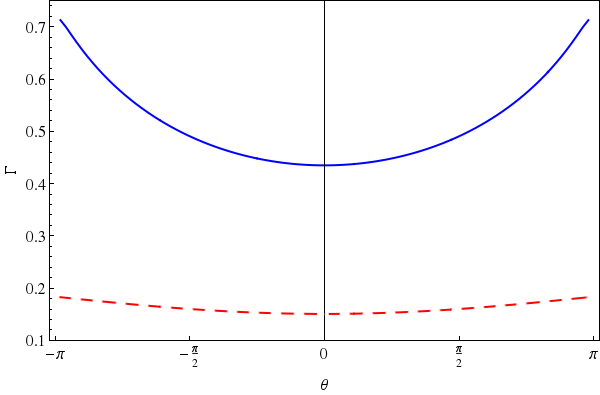}
\caption{ The $\theta$-dependence of the masses (left panel) and widths 
(right panel) of the $\sigma$ (blue, full) and the $\rho$ (red, dashed).
\label{fig:sigmarho}}
\end{figure}

When isospin breaking is
neglected, $m_u=m_d$, the vacuum is not shifted, and we can use the usual ChPT 
Lagrangian and amplitudes directly. All the $\theta$-dependence of physical observables enters 
through Eq.~\eqref{eq:LOmass} which finally leads to the mass and width of the $\sigma$ and the $\rho$ 
as a function of $\theta$, shown in Fig.~\ref{fig:sigmarho}. The $\theta$-dependence of the $\sigma$ mass
is stronger than the one of the $\rho$ mass since the former is in a $S$-wave
while the latter is in a $P$-wave. Also, both widths show a somewhat stronger
dependence on $\theta$ which is due to the enlarged phase space as the pion
mass decreases from its physical value when $|\theta|$ increases from 0.


\section{Summary and outlook}
\label{sec:sum}

In this paper, we have studied the $\theta$-dependence of the lightest
resonances in QCD. For that, we have derived the charged and neutral pion masses and the
pion-pion scattering amplitude at NLO in
the $\theta$-vacuum. We found that the NLO contributions proportional to $l_3$ and $l_7$
and entering via the pion mass formula and $\pi\pi$-contact terms cancel each other exactly.
The $\sigma$ and the $\rho$ have been obtained from a unitarization of this 
amplitude using the so-called (modified) inverse amplitude method. This automatically 
generates  $\theta$-dependent masses and widths of these resonances. Although the 
pion mass vanishes at $\theta = \pi$ at LO, no dramatic effects on the masses and 
widths of the  $\sigma$ and the $\rho$ were found. However, it still remains to be 
seen how such modifications change the properties of nuclei, as the nuclear binding 
is fine tuned, and thus more sensitive to such parameter variations.

\subsection*{Acknowledgements}

We are grateful to Christoph Hanhart and Jos\'e Antonio Oller for useful
discussions. This work is supported in part by the DFG and the NSFC
through funds provided to the Sino-German CRC~110 ``Symmetries and the Emergence of
Structure in QCD'' (NSFC Grant No. 11261130311). F.-K. G. acknowledges partial
support from the NSFC (Grant No. 11165005).

\bigskip

\begin{appendix}

\section{ Vacuum alignment at NLO }
\label{app:alignment}
\renewcommand{\theequation}{\thesection.\arabic{equation}}
\setcounter{equation}{0}

In this appendix, we will discuss the vacuum alignment at NLO induced by the
presence of the counterterms ($l_i$-terms) and the one-loop contribution. We
will calculate the vacuum alignment perturbatively.

Because the shift of the LO vacuum alignment given in Eq.~\eqref{eq:phi} is
caused by the NLO terms in the chiral expansion, we assume that the angle
$\varphi$ in $U_0 = \text{diag}\{ e^{i\varphi}, e^{-i\varphi} \}$ can be split
into
\begin{equation}
  \varphi = \varphi_0 + \alpha\, \varphi_1 \,
  \label{eq:phisplit}
\end{equation}
with $\varphi_0$ determined by aligning the vacuum at LO, and $\alpha\,
\varphi_1$ is the shift coming from the $\order{p^4}$ contribution to the
vacuum energy. Here, $\alpha$ is a chiral scaling factor to make explicit that
$\alpha\, \varphi_1$ is one order higher than $\varphi_0$ in the chiral
expansion.\footnote{The introduction of the scaling factor is only for
convenience. It will be set to $\alpha=1$ after $\varphi_1$ is calculated.}
The vacuum energy density up to NLO is given by
\begin{eqnarray}
  e_\text{vac} \al=\al  - \frac{F^2}{4}  \Tr{ \chi^\dag U_0 + \chi U_0^\dag }
  - \frac{l_3}{16} \Tr{   \chi^\dag U_0 + \chi U_0^\dag
  }^2 + \frac{l_7}{16} \Tr{\chi^\dag U_0 - \chi U_0^\dag }^2 \nonumber\\
  \al\al - \frac{h_1+h_3}{4} \Tr{\chi^\dag \chi} - \frac{h_1-h_3}{2}
  \text{Re}\,(\text{det}\chi) +
  e_\text{vac}^{(\text{1-loop})}  \, ,
  \label{eq:evac}
\end{eqnarray}
where  we have neglected those $h_i$-terms which are
independent of $U_0$, and $e_\text{vac}^{(\text{1-loop})}$ is the 1-loop
effective potential whose
 explicit expression 
 is~\cite{Guo:2015oxa} (see also Ref.~\cite{Bernard:2012ci} expanded up to
 $\theta^4$)
 \begin{equation}
   e_\text{vac}^{(\text{1-loop})} = 3 \mathring{M}^4(\theta) \left[ -
   \frac{\lambda}{2} + \frac1{128\pi^2} \left( 1 -2 \ln
   \frac{\mathring{M}^2(\theta)}{\mu^2} \right) \right]\,.
   \label{eq:evecloop}
 \end{equation}

We may decompose the vacuum energy density into the LO and NLO contributions
with the NLO one including all terms proportional to $\alpha$
\begin{equation}
 e_\text{vac} =  e_\text{vac} ^{(2)} + \alpha\,  e_\text{vac}^{(4)} \, .
\end{equation}
Substituting $U_0 =
\text{diag}\{ e^{i\varphi}, e^{-i\varphi} \}$ and Eq.~\eqref{eq:phisplit} into
Eq.~\eqref{eq:evac}, we get
\begin{eqnarray}
  e_\text{vac}^{(2)} \al=\al - F^2 M_\varphi^2, \qquad
  M_\varphi^2 = 2 B \bar m \left( \cos\frac{\theta}{2}
  \cos\varphi_0 - \epsilon \sin\frac{\theta}{2} \sin\varphi_0 \right)\,,
  \nonumber\\
  e_\text{vac}^{(4)} \al=\al \varphi_1 \frac{\partial\, e_\text{vac}^{(2)}}
  {\partial\, \varphi_0}
  + e_\text{vac}^{(\text{1-loop})}  - \frac{l_3}{F^4}\left( e_\text{vac}^{(2)}
  \right)^2 
  - 4 l_7 B^2 \bar m^2 \left( \sin\frac{\theta}{2} \cos\varphi_0 +
  \epsilon \cos\frac{\theta}{2} \sin\varphi_0 \right)^2\,,
\end{eqnarray}
where we have neglected the terms independent of $\varphi$.
The LO vacuum alignment is obtained by minimizing $e_\text{vac}^{(2)}$,
\begin{equation}
  \frac{\partial\, e_\text{vac}^{(2)}} {\partial\, \varphi_0} = 0\, ,
  \label{eq:vacuumLO}
\end{equation}
whose solution is given by $\varphi_0 = \bar \varphi_0$ with
\begin{equation}
  \bar \varphi_0 = \arctan \left( - \epsilon \tan\frac{\theta}{2} \right) \, .
  \label{eq:phi0}
\end{equation}
With this value of $\varphi_0$, the LO vacuum energy density, normalized to 0 at
$\theta=0$, is~\cite{Brower:2003yx}
\begin{equation}
  e_\text{vac}^{(2)} = F^2 \left[ M^2(0) - \mathring{M}^2(\theta) \right]\,,
\end{equation}
where $M^2(0) = 2 B\bar m$, and $\mathring{M}^2(\theta)$ is given in
Eq.~\eqref{eq:mpiLO}.
The NLO vacuum energy density is
\begin{equation}
  e_\text{vac}^{(4)} = e_\text{vac}^{(\text{1-loop})} - \mathring{M}^4(\theta)
  \left\{ l_3 + l_7 \left[\frac{(1-\epsilon^2) \tan(\theta/2)}{ 1 +
  \epsilon^2\tan^2(\theta/2) } \right]^2 \right\} \,.
\end{equation}

\begin{figure}[t]
\centering
\includegraphics[width=0.69\linewidth]{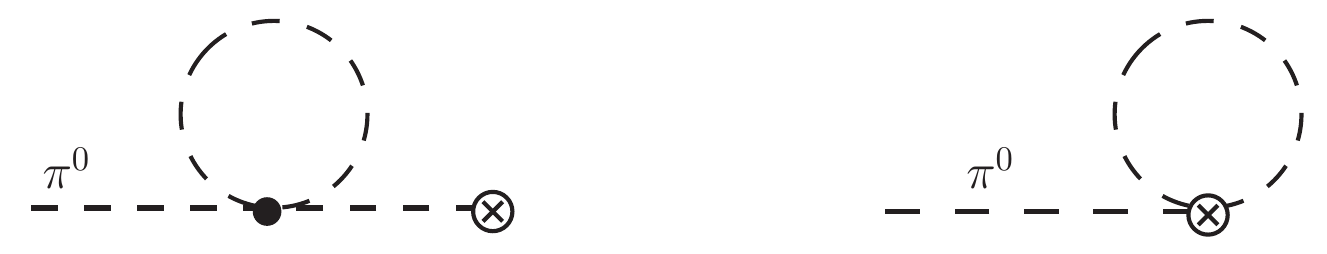}
\caption{ One-loop diagrams for producing a neutral pion from the vacuum. Here
$\otimes$ and the black dot denote CP violating and conserving vertices from the
LO Lagrangian, respectively.
\label{fig:onepion}}
\end{figure}

The perturbation $\varphi_1$ due to the NLO terms
is then determined by
\begin{equation}
  \left. \frac{\partial\, e_\text{vac}^{(4)}} {\partial\, \varphi_0}
  \right|_{\varphi_0 = \bar\varphi_0} = 0\,.
  \label{eq:vacuumNLO}
\end{equation}
From Eq.~\eqref{eq:vacuumLO}, it is easy to see that the $l_3$ term does not
have any effect. The vacuum alignment is equivalent to removing the
tadpole of the neutral pion which causes vacuum
instability~\cite{Dashen:1970et} (see also, e.g.,
Refs.~\cite{Crewther:1979pi,Mereghetti:2010tp,Bsaisou:2012rg,Bsaisou:2014oka}).
In fact, with $\varphi_0 = \bar\varphi_0$, the SU(2) LO chiral Lagrangian does
not have any term with odd number of pions because such a term is always
proportional to
\begin{equation}
  \cos\frac{\theta}{2} \sin\varphi_0 + \epsilon \sin\frac{\theta}{2}
  \cos\varphi_0 \propto \frac{\partial\, e_\text{vac}^{(2)}} {\partial\,
  \varphi_0} \, .
\end{equation}
This implies that the one-loop diagrams for producing a $\pi^0$ from the vacuum
as shown in Fig.~\ref{fig:onepion} have a vanishing amplitude.
Thus, we have $\left(\partial\,e_\text{vac}^{(\text{1-loop})}/\partial\,\varphi_0
\right)_{\varphi_0 = \bar\varphi_0} = 0$. This can be checked explicitly with
Eq.~\eqref{eq:evecloop} noticing that $\mathring M^2(\theta) = \left.
M_\varphi^2 \right|_{\varphi_0=\bar \varphi_0}$.
Therefore, Eq.~\eqref{eq:vacuumNLO} leads to a solution $\varphi_1 = \bar\varphi_1$ with
\begin{equation}
  \bar\varphi_1 = \frac{4 l_7 B\bar m}{F^2} \epsilon(1-\epsilon^2) \frac{
  \tan(\theta/2) \sec(\theta/2) }{\left[1 + \epsilon^2 \tan^2(\theta/2)
  \right]^{3/2} }\, .
\end{equation}
This is the NLO perturbation to the LO vacuum alignment.
Expanding it around $\theta = 0$,  we reproduce the result derived in
Ref.~\cite{Bsaisou:2014oka}
\begin{equation}
\bar\varphi_1 = \frac{2 l_7 B\bar m}{F^2} \epsilon(1-\epsilon^2) \theta +
\order{\theta^2}\, .
\label{eq:phi1}
\end{equation}

This perturbation produces a CP violating three-pion
vertex~\cite{Bsaisou:2012rg,Bsaisou:2014oka} by substituting $U_0$ with $\varphi  =
\bar\varphi_0 + \bar\varphi_1$ into the LO Lagrangian, which turns out to be of
$\order{p^4}$. Such a vertex contributes to the $\pi\pi$
scattering from $\order{p^6}$ and to the pion mass only starting at $\order{p^8}$.
Furthermore, terms with even number of pions are CP conserving and receive
contributions with an even power of $\bar\varphi_1$, so that the
$\bar\varphi_1$-induced terms in the Lagrangian also start from $\order{p^6}$.
Therefore, it is safe to make the vacuum alignment at LO for our calculation. It
is for the same reason that the topological susceptibility up to NLO in the
chiral expansion calculated in Ref.~\cite{Bernard:2012ci} agrees with that in
Ref.~\cite{Mao:2009sy}, where the vacuum was aligned by minimizing the vacuum
energy at LO and NLO, respectively.

\end{appendix}

\end{document}